\documentclass[cameraready]{Interspeech}

\usepackage{amsmath,amssymb,amsfonts}
\usepackage{algorithmic}
\usepackage{graphicx}
\usepackage{hyperref}
\usepackage{textcomp}
\usepackage{xcolor}
\usepackage{kotex}
\usepackage{pifont}
\usepackage{booktabs}
\usepackage{multirow}

\title{Evaluating Hallucinations in Audio-Visual Multimodal LLMs\\with Spoken Queries under Diverse Acoustic Conditions}

\author[affiliation={1}]{Hansol}{Park}
\author[affiliation={1}]{Hoseong}{Ahn}
\author[affiliation={2}]{Junwon}{Moon}
\author[affiliation={2}]{Yejin}{Lee}
\author[affiliation={1,2}]{Kyuhong}{Shim}

\address{
    $^1$ Department of Intelligent Software, Sungkyunkwan University \\
    $^2$ Department of Computer Science and Engineering, Sungkyunkwan University
}

\email{ \{firstri, hoseong8115, mppn98, yj.lee, khshim\}@skku.edu }


\usepackage{comment}

\begin{document}
\maketitle

\begin{abstract}
Hallucinations in multimodal models have been extensively studied using benchmarks that probe reliability in image-text query settings. 
However, the effect of spoken queries on multimodal hallucinations remains largely unexplored, despite the growing role of voice interfaces. 
In this paper, we introduce a systematic pipeline that converts existing multimodal hallucination benchmarks into spoken-query versions while preserving the original tasks and labels. 
We instantiate this pipeline on RePOPE and release RePOPE-Spk, where all queries are provided as spoken audio under diverse input conditions. 
Experimental results show that hallucinations escalate when queries are spoken rather than written: error rates increase by 3--6\% with clean speech and by up to 30\% under environmental noise. 
Furthermore, many-shot prompting and chain-of-thought reasoning provide only partial mitigation.
Our findings motivate new directions for building reliable voice interface systems and evaluations.
\end{abstract}

\section{Introduction}\label{sec:intro}

Voice-driven interaction is rapidly becoming a primary interface for artificial intelligence (AI) systems, powering mobile assistants, smart devices, and emerging AR/VR platforms.
As these technologies move toward more immersive and hands-free experiences, spoken language will increasingly replace typed input in human-AI interaction.
Multimodal large language models (MLLMs), which can process text, images, and audio, play a central role in this shift~\cite{xu2025qwen3,li2025baichuan}.
By leveraging heterogeneous multimodal signals, these models can understand complex contexts and perform grounded reasoning that was previously unattainable with single-modality systems.

A major concern is hallucination, where models produce outputs inconsistent with the given context or contradict factual evidence.
Hallucinations occur more frequently in MLLMs than in single-modality models, largely due to cross-modal interactions~\cite{leng2024curse,jung2025avcd}.
For instance, a model may claim the existence of objects or events not present in the input, or arbitrarily add details that are not provided.
Hallucinations in vision–language models have been extensively studied through benchmarks such as POPE~\cite{li2023pope}, CHAIR~\cite{rohrbach2018chair}, and AMBER~\cite{wang2023amber}.
However, scenarios involving spoken queries remain underexplored: when audio is included, it is typically treated as synchronized context (e.g., paired with video), rather than as an independent query modality~\cite{sungbin2025avhbench,jung2025avcd}.
This gap is important because voice interfaces introduce acoustic variability that can distort the query content.
As a result, evaluations based only on written queries may overestimate reliability in real-world voice interaction.

\begin{figure}[t!]
     \centering\includegraphics[width=1.0\linewidth]{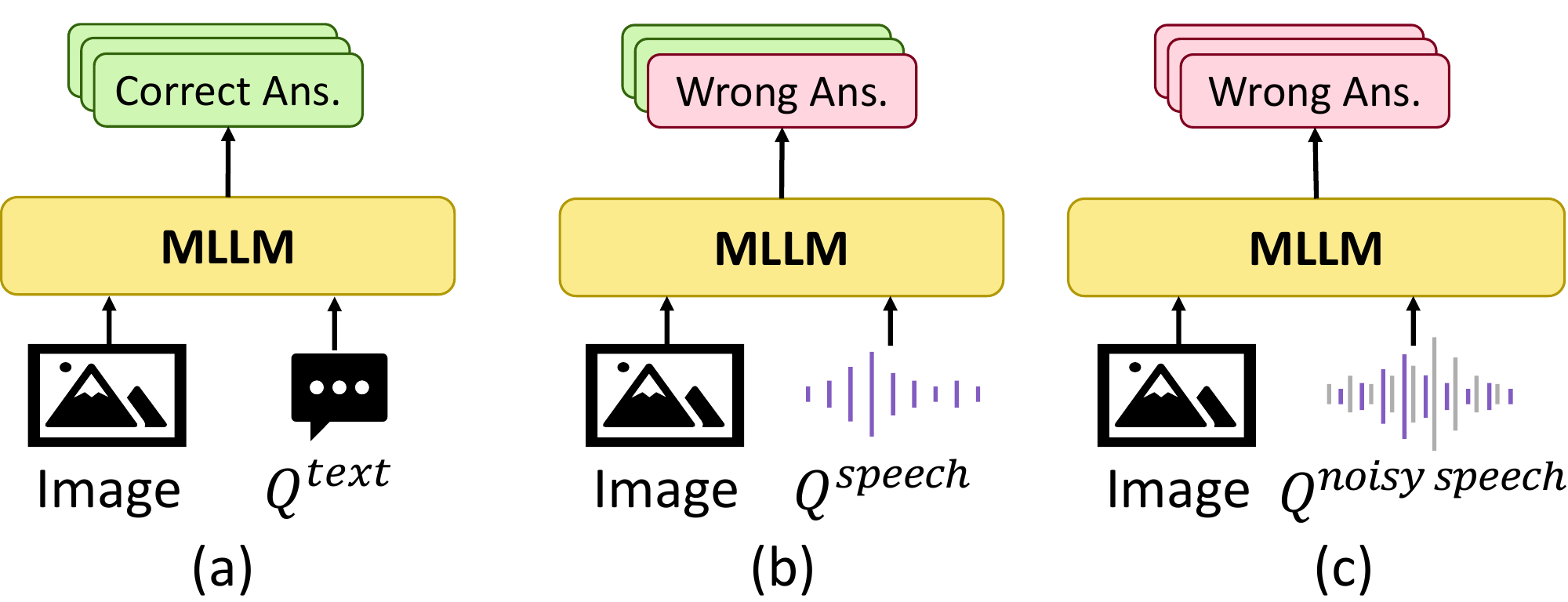}
     \vspace{-0.3cm}
     \caption{Spoken queries amplify multimodal hallucinations: (a) image–text queries yield correct answers, (b) replacing the text query with semantically equivalent speech increases hallucinations, and (c) adding noise to the spoken query further exacerbates them.}
     \label{fig:intro}
     \vspace{-0.2cm}
 \end{figure}

Motivated by this gap, we address two empirical questions: 1) How does replacing a text query with a spoken query change hallucination behavior in audio-visual MLLMs? 2) Do common reasoning and prompting techniques remain effective under spoken-query conditions?
We hypothesize that replacing text queries with speech systematically increases hallucinations, and that acoustic variability (e.g., environmental noise) amplifies this effect (see Figure~\ref{fig:intro}).
To study these questions and evaluate the hypothesis, we propose a benchmark conversion pipeline that transforms existing multimodal hallucination benchmarks into spoken-query counterparts.
Because our conversion preserves the original task definition and labels, it enables apples-to-apples comparisons that isolate the effect of query modality and acoustic conditions on hallucination, rather than conflating it with dataset shifts.

As a concrete instantiation of this pipeline, we introduce RePOPE-Spk, a spoken-query extension of the RePOPE~\cite{neuhaus2025repope} benchmark, where all queries are provided as spoken audio under diverse acoustic and input conditions.
Using RePOPE-Spk, we conduct controlled experiments on both proprietary and open-source MLLMs, revealing consistent weaknesses in current models.
Our findings highlight an overlooked challenge at the intersection of speech and multimodal reasoning, and suggest new research directions for reliable voice AI.

\vspace{0.1cm}
\noindent Our contributions are summarized as follows.
\begin{itemize}
    \item We propose a systematic conversion pipeline for transforming existing benchmarks into spoken-query versions.
    \item We analyze the impact of noise, input order, and query length on hallucination rates across representative MLLMs.
    \item We establish training-free baselines (e.g., many-shot prompting and chain-of-thought reasoning) and characterize their efficiency. Our result underscores the intrinsic difficulty of robust voice-based multimodal interaction.
\end{itemize}
\begin{figure*}[t]
     \centering\includegraphics[width=1.0\linewidth]{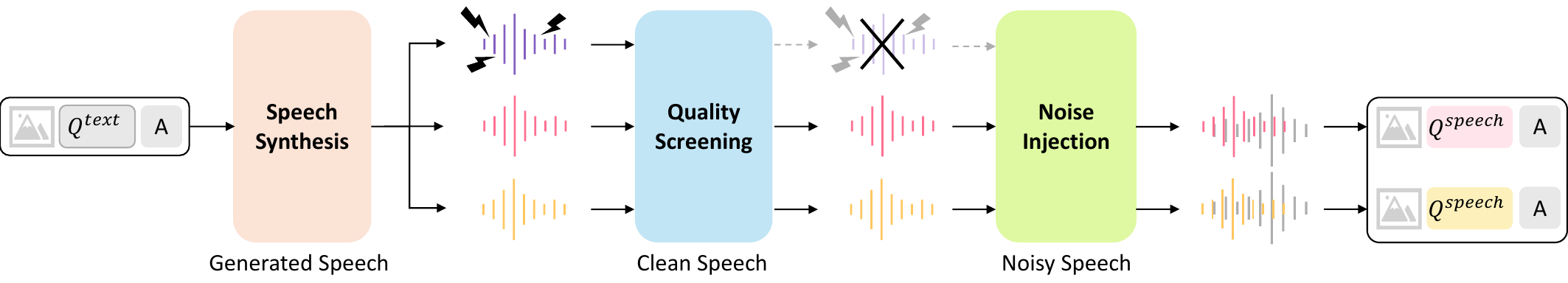}
     \vspace{-0.2cm}
     \caption{Benchmark conversion pipeline. The proposed framework converts an image-text dataset into an image-spoken query dataset using controlled perturbations to generate diverse and challenging spoken queries.}
     \label{fig:method}
\end{figure*}

\section{Related Work}\label{sec:related}

\subsection{Hallucination in Audio Multimodal Models}\label{ssec:related_multimodal}
As MLLMs expand beyond text and vision, hallucinations in the audio modality are becoming increasingly important.
CMM~\cite{leng2024curse} demonstrated that adding audio, alongside language and vision, can amplify spurious correlations and mislead model predictions across modalities.
In the audio–visual domain, AVHBench~\cite{sungbin2025avhbench} and AVCD~\cite{jung2025avcd} probe cross-modal hallucinations, such as falsely inferring non-existent visual events from audio or vice versa.
Similarly, recent studies~\cite{kuan24understanding,kuan2025can} report that audio-language models frequently hallucinate non-existent sounds or misinterpret temporal and attribute details.
On the other hand, several mitigation strategies have been proposed.
LISTEN~\cite{kuan2025teaching} and AAD~\cite{hsu2025reducing} leverage negative samples or contrastive decoding to reduce such errors.
However, these works primarily focus on perception-driven audio or audio-visual grounding, without examining spoken queries.

\subsection{Spoken Queries in LLMs}\label{ssec:related_spoken}
A line of research investigates how LLMs handle spoken queries, though these studies primarily remain text-oriented.
VoiceBench~\cite{chen2024voicebench} and SpeechIQ~\cite{wan2025speechiq} evaluate the ability of multimodal LLMs to comprehend the spoken instructions.
New datasets such as SpokenNativQA~\cite{alam2025spokennativqa} provide multilingual spoken queries with diverse accents and prosody, while CCFQA~\cite{du2025ccfqa} evaluates factual consistency in speech–text question-answering across languages.
These resources highlight the potential of spoken input to stress-test MLLMs, yet they focus on factuality and coverage rather than hallucination.
Moreover, robustness to acoustic variability, such as environmental noise or input order, has received little attention.

\section{Text-to-Spoken Query Conversion}\label{sec:method}

\subsection{Benchmark Conversion Pipeline}
We propose a conversion pipeline that transforms text-query multimodal hallucination benchmarks into spoken-query counterparts (see Figure~\ref{fig:method}).
Given a VQA-style sample $(I, Q^{\text{text}}, A)$ consisting of an image $I$, a text query $Q^{\text{text}}$, and a ground-truth answer $A$, the pipeline generates one or more spoken queries $Q^{\text{speech}}$ under controlled acoustic conditions. 
The resulting samples $(I, Q^{\text{speech}}, A)$ enable direct comparison between written- and spoken-query evaluation.
By varying the conversion configuration (e.g., noise conditions), the pipeline produces multiple variants of the same benchmark, which facilitates analysis of acoustic factors.
The pipeline consists of three steps: {(i) speech synthesis}, {(ii) quality screening}, and {(iii) noise injection}. 

\subsubsection{Speech Synthesis}
We first generate clean spoken queries from original texts using a high-quality TTS model (e.g., Parler-TTS~\cite{lyth2024parler} and Edge-TTS~\footnote{\texttt{https://github.com/rany2/edge-tts}}).
We utilize a set of consistent speaking styles to minimize confounds from speaker variability and recording conditions.
This design isolates the impact of the query modality and the controlled perturbations introduced in the next step.

\subsubsection{Quality Screening}
Although modern TTS systems are generally reliable, occasional synthesis errors may introduce unintended semantic changes.
To ensure that each spoken query faithfully preserves the original content, we perform automatic screening using ASR~\cite{wang2025tens,zhao2026towards}. 
Specifically, we transcribe each synthesized utterance with Whisper~\cite{radford2023whisper}, compute WER against the original query text, and discard samples with WER exceeding a predefined threshold. 
For rejected samples, we resynthesize with different random seeds up to a fixed maximum number of attempts and retain the first utterance that satisfies the WER criterion. 
In practice, this screening rejects only a small fraction of samples.

\subsubsection{Noise Injection}
To evaluate robustness under realistic acoustic conditions, we mix spoken queries with environmental noise from ESC-50~\cite{piczak2015dataset}, which includes diverse sound classes (e.g., human, animal, indoor, outdoor, and natural sounds). 
We generate noisy speech at SNR levels of 0~dB and 5~dB following common practice~\cite{zhu2001noise}.
Unless noted otherwise, noise segments are randomly sampled and mixed with the utterance to match its duration.
All stochastic components (e.g., TTS sampling, noise segment selection, and mixing) are controlled by fixed random seeds for reproducibility.

\setlength{\tabcolsep}{10.5pt}
\begin{table*}[t!]
    \centering
    \caption{Vulnerability of multimodal LLMs to hallucination under different input conditions. 
    The inputs are either text, clean audio, or noisy audio with added noise at 0 dB and 5 dB SNR. 
    A, R, P, and F1 denote accuracy, recall, precision, and F1 score, respectively.}
    \vspace{-0.1cm}
    \resizebox{1.0\linewidth}{!}{
    \begin{tabular}{cc|cccc|cccc|cccc}
        \toprule[1.5pt]
        \multirow{2}{*}{Model} & \multirow{2}{*}{Input/Noise} & \multicolumn{4}{c|}{Adversarial (\%)} & \multicolumn{4}{c|}{Popular (\%)}         & \multicolumn{4}{c}{Random (\%)} \\
                   &    & A & R & P & F1 & A & R & P & F1 & A & R & P & F1 \\
        \midrule[1.5pt]
        Gemini & Text & 92.6 & 88.4 & 95.9 & 92.0 & 94.3 & 91.9 & 95.5 & 93.7 & 96.8 & 96.2 & 96.2 & 96.2\\
        -1.5-Flash          & Clean audio & 90.9 & 87.4 & 92.7 & 90.2 & 91.6 & 88.7 & 92.5 & 90.6 & 92.5 & 89.1 & 93.5 & 91.2\\
        \midrule
            \multirow{6}{*}{+5 dB noise} 
                         & Animals  & 83.1 & 93.9 & 74.6 & 83.1 & 83.5 & 93.9 & 74.8 & 83.2 & 83.9 & 94.6 & 74.0 & 83.0\\
                         & Natural  & 79.6 & 95.2 & 69.7 & 80.5 & 79.8 & 93.7 & 70.1 & 80.2 & 78.2 & 94.7 & 66.8 & 78.4\\
                         & Human    & 83.1 & 94.3 & 74.2 & 83.0 & 83.1 & 93.4 & 74.4 & 82.8 & 83.0 & 95.3 & 72.4 & 82.3\\
                         & Interior & 83.9 & 93.6 & 75.5 & 83.6 & 84.4 & 93.7 & 76.2 & 84.0 & 84.0 & 94.1 & 74.4 & 83.1\\
                         & Exterior & 82.0 & 94.2 & 72.9 & 82.2 & 82.0 & 94.2 & 72.7 & 82.0 & 81.1 & 94.7 & 70.3 & 80.7\\
                         \cmidrule{2-14}
                         & Avg. & 82.3 & 94.2 & 73.4 & 82.5 & 82.6 & 93.8 & 73.6 & 82.4 & 82.0 & 94.7 & 71.6 & 81.5\\
         \midrule
            \multirow{6}{*}{+0 dB noise} 
                         & Animals  & 76.4 & 94.9 & 66.3 & 78.0 & 75.5 & 95.7 & 64.7 & 77.2 & 76.2 & 94.5 & 64.8 & 76.8\\
                         & Natural  & 71.2 & 94.9 & 61.0 & 74.3 & 71.0 & 94.8 & 60.7 & 74.0 & 69.6 & 95.4 & 58.1 & 72.3\\
                         & Human    & 76.3 & 94.1 & 65.8 & 77.4 & 76.3 & 95.1 & 65.9 & 77.8 & 74.3 & 94.7 & 61.9 & 74.9\\
                         & Interior & 78.2 & 94.2 & 67.9 & 78.9 & 77.7 & 94.0 & 67.6 & 78.7 & 77.7 & 94.8 & 66.0 & 77.8\\
                         & Exterior & 72.2 & 94.5 & 62.2 & 75.0 & 71.3 & 94.9 & 61.0 & 74.3 & 71.2 & 95.5 & 59.7 & 73.5\\
                         \cmidrule{2-14}
                         & Avg. & 74.9 & 94.5 & 64.6 & 76.7 & 74.4 & 94.9 & 64.0 & 76.4 & 73.8 & 95.0 & 62.1 & 75.1\\
        \midrule[1.5pt]
        \multirow{3}{*}{Gemma-3n}  & Text & 89.1 & 95.2 & 82.8 & 88.6 & 89.8 & 95.1 & 83.8 & 89.0 & 91.6 & 95.6 & 85.8 & 90.4\\
                        & Clean audio & 83.7 & 92.9 & 75.8 & 83.5 & 84.1 & 93.1 & 76.0 & 83.7 & 82.0 & 92.9 & 72.1 & 81.2\\
                        & ASR cascade & 88.3 & 94.5 & 81.9 & 87.7 & 89.6 & 94.0 & 84.1 & 88.8 & 92.6 & 95.1 & 88.1 & 91.5\\
        \midrule
            \multirow{6}{*}{+5 dB noise} 
                         & Animals & 69.9 & 95.7 & 60.0 & 73.7 & 67.8 & 95.7 & 58.0 & 72.2 & 66.0 & 94.7 & 55.4 & 69.9\\
                         & Natural & 66.6 & 95.4 & 57.4 & 71.6 & 65.3 & 95.9 & 56.1 & 70.7 & 62.8 & 94.7 & 53.0 & 68.0\\
                         & Human & 68.7 & 95.2 & 59.1 & 72.9 & 67.2 & 94.8 & 57.6 & 71.7 & 64.3 & 95.3 & 54.2 & 69.1\\
                         & Interior & 70.7 & 94.2 & 60.9 & 74.0 & 70.1 & 94.8 & 60.0 & 73.5 & 68.6 & 94.7 & 57.5 & 71.6\\
                         & Exterior & 67.7 & 95.5 & 58.2 & 72.3 & 65.5 & 94.6 & 56.3 & 70.6 & 63.6 & 95.1 & 53.7 & 68.6\\
                         \cmidrule{2-14}
                         & Avg. & 68.7 & 95.2 & 59.1 & 72.9 & 67.1 & 95.1 & 57.6 & 71.7 & 65.1 & 94.9 & 54.8 & 69.4\\
         \midrule
            \multirow{6}{*}{+0 dB noise} 
                         & Animals & 62.9 & 96.7 & 54.6 & 69.8 & 60.0 & 96.4 & 52.3 & 67.8 & 59.1 & 95.9 & 50.5 & 66.2\\
                         & Natural & 60.4 & 97.1 & 52.8 & 68.4 & 59.8 & 97.2 & 52.2 & 67.9 & 56.6 & 95.9 & 49.0 & 64.8\\
                         & Human & 62.7 & 95.8 & 54.4 & 69.4 & 61.2 & 95.8 & 53.1 & 68.4 & 57.9 & 96.2 & 49.8 & 65.6\\
                         & Interior & 65.0 & 95.4 & 56.1 & 70.7 & 65.4 & 96.1 & 56.1 & 70.9 & 63.0 & 95.2 & 53.2 & 68.2\\
                         & Exterior & 59.4 & 96.8 & 52.2 & 67.8 & 56.9 & 96.6 & 50.4 & 66.2 & 56.2 & 96.0 & 48.8 & 64.7\\
                         \cmidrule{2-14}
                         & Avg. & 62.1 & 96.4 & 54.0 & 69.2 & 60.7 & 96.4 & 52.8 & 68.2 & 58.6 & 95.8 & 50.3 & 65.9\\
        \bottomrule[1.5pt]
    \end{tabular}}
    \label{tab:exp_type_noise}
\end{table*}
\setlength{\tabcolsep}{6pt}

\subsection{Instantiating the Pipeline: RePOPE-Spk}

We instantiate the proposed pipeline on RePOPE~\cite{neuhaus2025repope}, a widely used vision--language benchmark for object hallucination, where models incorrectly claim the presence or attributes of objects not grounded in the image. 
RePOPE refines POPE~\cite{li2023pope} by correcting annotation inconsistencies and provides three subsets: \textit{Adversarial} (2,684 QAs), \textit{Popular} (2,727 QAs), and \textit{Random} (2,774 QAs). 
We refer to the new benchmark produced by our pipeline as \textbf{RePOPE-Spk}.
The benchmark preserves the original images, questions, answers, and subset definitions, and converts only the text queries into spoken ones.
This supports direct and controlled comparisons between text-query and spoken-query evaluation. 

\section{Experiments}\label{sec:experiment}

\subsection{Setup}\label{ssec:experiment_datasets}
We evaluate two representative MLLMs, Gemini~\cite{team2024gemini} (proprietary) and Gemma~\cite{team2025gemma} (open-source), that support unified text, image, and audio inputs.
To generate spoken queries, we use Parler-TTS~\cite{lyth2024parler} with a single consistent voice profile to control for speaker variability.
Using RePOPE-Spk, we measure hallucination behavior under spoken queries with diverse acoustic conditions.
For Tables~\ref{tab:exp_type_noise} and~\ref{tab:exp_order}, we report performance in all three subsets following the original benchmark~\cite{li2023pope}.
Other experiments are conducted on \textit{Adversarial} subset with Natural-category noise conditions.
Following the original benchmark, we report accuracy, recall, precision, and F1 score.

\subsection{Effect of Noise in Spoken Query}
Table~\ref{tab:exp_type_noise} summarizes the impact of modality conversion and noise injection.
We observe three consistent trends.
First, spoken queries degrade performance relative to text queries.
Switching from text to clean audio lowers accuracy and F1 score across both models, confirming that speech introduces an inherent loss.
Second, stronger noise worsens hallucinations.
Performance degrades substantially as SNR drops from 5 dB to 0 dB, particularly in precision, indicating that models generate more false positives under noisy conditions.
Third, although Gemini achieves higher absolute scores than Gemma, both models exhibit comparable relative degradation as noise increases.
This suggests that the limitations arise from shared challenges in speech robustness rather than model-specific weaknesses.

\setlength{\tabcolsep}{12pt}
\begin{table}[t]
    \centering
    \caption{Effect of the order of input modalities. $I$ and $S$ denote the image and spoken query, respectively.}
    \vspace{-0.15cm}
    \resizebox{1.0\linewidth}{!}{
    \begin{tabular}{cc|cccc}
        \toprule[1.2pt]
        Model & Order & A & R & P & F1 \\
        \midrule[1.2pt]
        \multirow{2}{*}{Gemini} & $I\rightarrow S$ & 90.9 &  87.4 & 92.7 & 90.2\\
                                & $S\rightarrow I$ & 90.1 &  88.5 & 89.2 & 88.8\\
        \midrule
        \multirow{2}{*}{Gemma}  & $I\rightarrow S$ & 83.7 & 92.9 & 75.8 & 83.5\\
                                & $S\rightarrow I$ & 69.0 &  82.0 & 61.2 & 70.1\\
        \bottomrule[1.2pt]
    \end{tabular}}
    \vspace{-0.5cm}
    \label{tab:exp_order}
\end{table}
\setlength{\tabcolsep}{6pt}

\setlength{\tabcolsep}{14pt}
\begin{table}[t]
    \centering
    \caption{Effect of spoken query length (silence padding).}
    \vspace{-0.15cm}
    \resizebox{1.0\linewidth}{!}{
    \begin{tabular}{l|cccc}
        \toprule[1.2pt]
        Input & A & R & P & F1 \\
        \midrule[1.2pt]
        \multicolumn{5}{c}{Gemini-1.5-Flash} \\
        \midrule
        Original (clean) & 90.9 &  87.4 & 92.7 & 90.2\\
        \midrule
        Original (+0 dB) & 71.2 & 94.9 & 61.0 & 74.3 \\
        5s (+0 dB) & 79.7 & 92.4 & 70.6 & 80.1 \\
        10s (+0 dB) & 82.9 & 94.0 & 74.2 & \textbf{82.9} \\
        \midrule[1.2pt]
        \multicolumn{5}{c}{Gemma-3n} \\
        \midrule
        Original (clean) & 83.7 & 92.9 & 75.8 & 83.5\\
        \midrule
        Original (+0 dB) & 60.4 & 97.1 & 52.8 & 68.4 \\
        5s (+0 dB) & 71.1 &  95.5 & 61.1 & 74.5\\
        10s (+0 dB) & 74.7 &  95.1 & 64.5 & \textbf{76.9}\\
        \bottomrule[1.2pt]
    \end{tabular}}
    \label{tab:exp_length}
\end{table}
\setlength{\tabcolsep}{6pt}

\setlength{\tabcolsep}{13pt}
\begin{table}[t]
    \centering
    \caption{Many-shot prompting on Gemini-1.5-Flash.}
    \vspace{-0.15cm}
    \resizebox{1.0\linewidth}{!}{
    \begin{tabular}{cc|cccc}
        \toprule[1.2pt]
        Noise & \#Shots & A & R & P & F1 \\
        \midrule[1.2pt]
        \multirow{3}{*}{Clean}  & 1 & 71.8 & 92.7 &	62.2 &	74.4\\
                                & 5 & 91.2 & 88.3 &	91.4 &	\textbf{89.8}\\
                                & 10 & 90.2	& 89.7 & 88.3 & 89.0\\
        \midrule
        \multirow{3}{*}{+5 dB}  & 1 & 59.0 & 94.0 &	52.0 & 66.9\\
                                & 5 & 86.1 & 89.6 &	81.1 & \textbf{85.1}\\
                                & 10 & 84.7 & 90.5 & 78.2 &	83.9\\
        \midrule
        \multirow{3}{*}{+0 dB}  & 1 & 55.3 & 95.2 & 49.7 & 65.3\\
                                & 5 & 78.8 & 91.0 & 70.0 & \textbf{79.1}\\
                                & 10 & 78.1 & 92.3 & 68.5 & 78.7\\
        \bottomrule[1.2pt]
    \end{tabular}}
    \label{tab:exp_many_shot}
\end{table}
\setlength{\tabcolsep}{6pt}

\setlength{\tabcolsep}{11pt}
\begin{table}[t]
    \centering
    \caption{Chain-of-thought reasoning on Gemini-2.5-Flash.}
    \vspace{-0.15cm}
    \resizebox{1.0\linewidth}{!}{
    \begin{tabular}{l|cccc}
        \toprule[1.2pt]
        Type & A & R & P & F1 \\
        \midrule[1.2pt]
        Zero-shot (text)        & 94.6&94.7&93.3&94.0\\
        \midrule
        Zero-shot (+0 dB)       & 73.4 & 49.6 & 83.4 & 62.2\\
        + Step-by-Step          & 70.6 & 95.1 & 59.9 & 73.5\\
        + Visual-Description    & 77.6 & 83.0 & 71.1 & 76.6\\
        + Speech-Description    & 79.1 & 83.7 & 72.6 & 77.8\\
        + Cross-Modal           & 79.7 & 81.1 & 75.0 & \textbf{77.9}\\
        \bottomrule[1.2pt]
    \end{tabular}}
    \vspace{-0.2cm}
    \label{tab:exp_cot}
\end{table}
\setlength{\tabcolsep}{6pt}

\subsection{Effect of Input Modality Order}
Inspired by previous work~\cite{berglund2024reversal}, Table~\ref{tab:exp_order} compares two input orderings: image before spoken query ($I\rightarrow S$), and image after spoken query ($S\rightarrow I$).
Gemini is largely insensitive to input order, showing only minor performance changes.
In contrast, Gemma exhibits a severe degradation under ($S\rightarrow I$) ordering; the F1 score degrades from 83.5\% to 70.1\%.
This result indicates that Gemma is more sensitive to modality ordering, potentially reflecting differences in training data composition or weaknesses when modalities arrive in an unexpected sequence.

\subsection{Effect of Spoken Query Length}
Table~\ref{tab:exp_length} investigates the effect of increasing spoken query length by padding the input audio.
Extending the length of noisy queries from the original duration partially restores performance, which has not been reported in previous studies.
One plausible explanation is that longer audio provides additional acoustic context that stabilizes speech processing; however, the gains remain insufficient to eliminate degradation.

\begin{figure}[t!]
     \centering\includegraphics[width=1.0\linewidth]{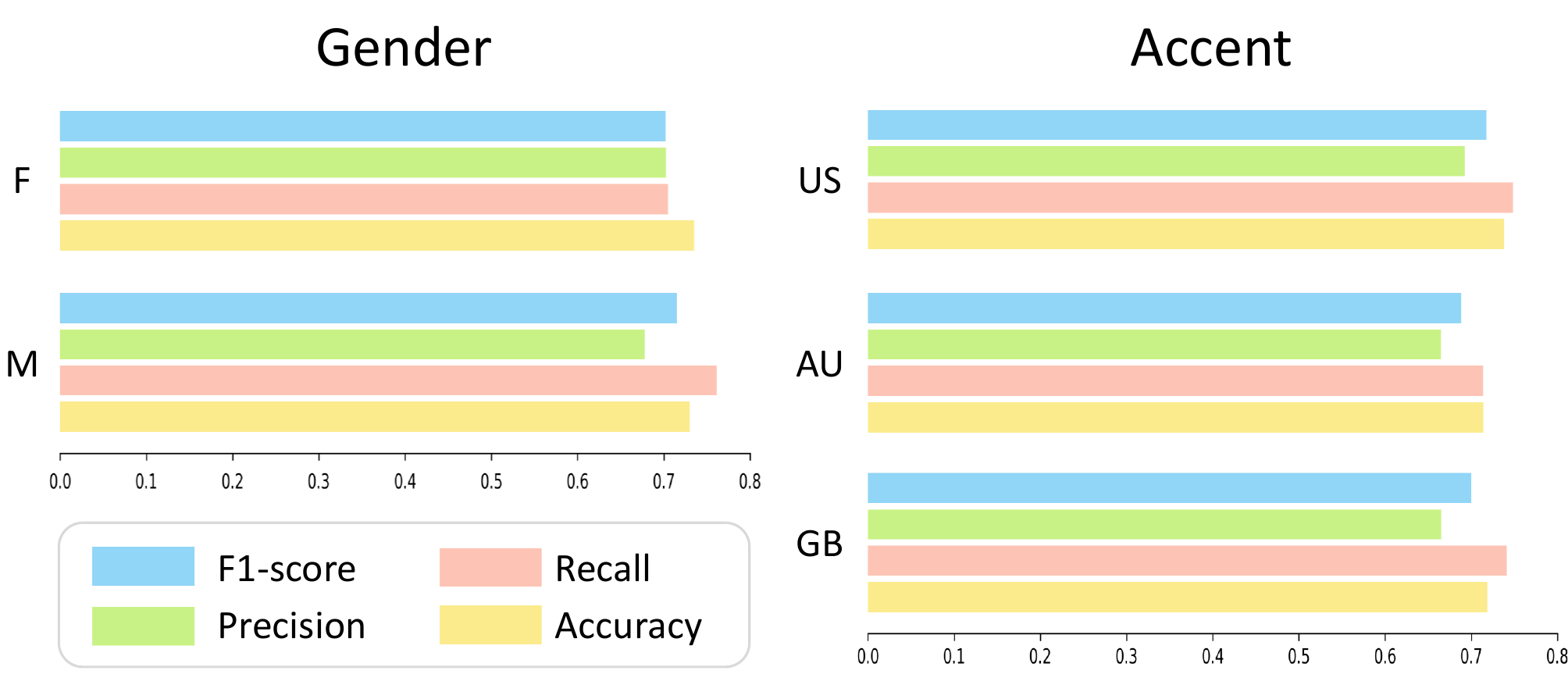}
     \vspace{-0.5cm}
     \caption{Effect of speaker profiles (gender and accent).}
     \label{fig:profile}
     \vspace{-0.2cm}
 \end{figure}

\subsection{Many-shot Prompting}
Table~\ref{tab:exp_many_shot} evaluates the effect of in-context learning.
Increasing the number of shots from 1 to 5 shows clear gains across all noise levels, including clean audio.
However, performance declines beyond 5 shots, contrasting with text-only LLMs, where more examples typically help~\cite{agarwal2024many}.
The observed plateau suggests that speech-based prompting may be less scalable and does not benefit from the straightforward adoption of text-based prompting techniques~\cite{jiang2024manyshot}.

\subsection{Chain-of-Thought (CoT) Reasoning}
Table~\ref{tab:exp_cot} examines CoT~\cite{zhang2024mcot} prompting strategies using Gemini-2-5-Flash, which has stronger reasoning capabilities.
Notably, the performance of Gemini-2.5-Flash under noisy conditions falls below that of Gemini-1.5-Flash, underscoring the inherent difficulty of the task.
``Step-by-step'', the simplest CoT technique, improves robustness by encouraging explicit reasoning.
``Visual-Description'' directs the model to systematically investigate the image~\cite{ghosh2025visual}, whereas ``Speech-Description'' instructs the model to \textit{transcribe} the spoken query before answering~\cite{ma2025audiocot}.
Combining both (``Cross-Modal'') shows the best F1 score of 77.9\%, which significantly surpasses the baseline of 62.2\%.
Based on the performance gap between the above three prompting techniques, the gains primarily arise from explicitly addressing the speech modality.
Nevertheless, none of the strategies closes the gap to text-query performance.

\subsection{Effect of Speaker Profile}
We further evaluate robustness across eight different voice profiles provided by Edge-TTS.
We categorize these profiles by gender (male/female) and accent (US/AU/GB) (see Figure~\ref{fig:profile}).
Performance differences between male and female voices are negligible.
In contrast, changing accent leads to a measurable F1 decrease (1.8–3.0\%).
We suggest that the performance degradation is primarily driven by the use of spoken queries and their acoustic variability, rather than by speaker identity alone.

\subsection{Comparison with ASR-LLM Cascade}
Finally, we examine performance of ASR-transcribed queries at 0~dB SNR (see Table~\ref{tab:exp_type_noise}, ``ASR Cascade'').
While the ASR-LLM pipeline appears comparable to text input, this is largely an artifact of RePOPE's binary (yes/no) evaluation.
Even with incomplete or partially incorrect transcripts (i.e., high WER), models often answer correctly without faithfully understanding the query.
We observe severe semantic drift under heavy noise (e.g., ``Is there a dog in the image'' $\rightarrow$ ``It's very doggy, man''), yet the model may output the correct binary label.
This failure is consistent with the tendency of LLMs to avoid abstention~\cite{kalai2025language}.

We note that the ASR–LLM cascade system discards paralinguistic cues (e.g., prosody, emphasis, and speaking style) that can be essential for interpreting user intent in natural voice interaction~\cite{lin2024paralinguistics,yang2026para}.
Therefore, the system should be viewed as a temporary workaround rather than a solution; robust handling of spoken queries remains an open problem.
\section{Conclusion}\label{sec:conclusion}

In this work, we studied spoken-query hallucinations in MLLMs and introduced a systematic pipeline for converting text-query hallucination benchmarks into spoken-query counterparts under diverse acoustic conditions.
Using RePOPE-Spk, we showed that 1) spoken queries consistently degrade performance, 2) noise and modality order amplify hallucinations, and 3) strategies such as many-shot prompting and CoT offer limited gains.
Both proprietary and open-source models exhibited similar vulnerabilities, suggesting that the limitations are systemic rather than model-specific.
Our results highlight spoken-query robustness as an underexplored challenge and motivate new approaches for reliable speech-based interaction.


\newpage
\section{Generative AI Use Disclosure}
The authors used Claude (Anthropic) to assist with manuscript proofreading and writing style refinement.
All AI-assisted content was reviewed and verified by the authors.
\bibliographystyle{IEEEtran}
\bibliography{reference}

\end{document}